\documentclass[onecollarge]{svjour2}

\def\makeheadbox{{}}

\smartqed  % flush right qed marks, e.g. at end of proof
\usepackage[numbers,sort&compress]{natbib}
\usepackage{graphicx}
\usepackage{subfig}

\usepackage{multirow}
\usepackage{amsmath}
\usepackage{comment}
\usepackage{txfonts}
\usepackage{lineno}
\usepackage{tabularx}
\usepackage[latin1]{inputenc}
\usepackage{fancyhdr}
\graphicspath{{figures/}}
\newcommand{\AFTER}{\mbox{AFTER@LHC}}
\newcommand{\beq}{\begin{eqnarray}}
\newcommand{\eeq}{\end{eqnarray}}
\newcommand{\be}{\begin{eqnarray*}}
\newcommand{\ee}{\end{eqnarray*}}

\newcommand{\ie}{{\it i.e.}}
\newcommand{\eg}{{\it e.g.}}

\def\lsim{\raise0.3ex\hbox{$<$\kern-0.75em\raise-1.1ex\hbox{$\sim$}}}
\def\gsim{\raise0.3ex\hbox{$>$\kern-0.75em\raise-1.1ex\hbox{$\sim$}}}

\def\pp   {$pp$}

\def\sqrtsNN {\mbox{$\sqrt{s_{NN}}$}}

\def\pT      {\mbox{$p_{T}$}}
\def\kT     {\mbox{$k_{T}$}}

\def\beq     {\begin{equation}}
\def\eeq     {\end{equation}}

\def\pp   {$pp$}

\def\sqrtsNN {\mbox{$\sqrt{s_{NN}}$}}

\def\jpsi    {\mbox{$J/\psi$}}

\def \xF     {\mbox{$x_{F}$}}
\def \An     {\mbox{$A_{N}$}}
\newcommand{\gevc}{\mbox{$\mathrm{GeV/}c$}}
\newcommand{\gevcc}{\mbox{$\mathrm{GeV/}c^2$}}

\def\beq     {\begin{equation}}
\def\eeq     {\end{equation}}

\newcommand{\psip}{\mbox{$\psi'$}}

\newcommand{\ups}{\mbox{$\Upsilon$}}

\usepackage{ifpdf}
\ifpdf
\usepackage[pdftex]{hyperref}
\else
\usepackage[hypertex]{hyperref}
\fi

\hypersetup{
  pdftitle={Feasibility Studies for Single Transverse-Spin Asymmetry Measurements at a Fixed-Target Experiment Using the LHC Proton and
Lead Beams (AFTER@LHC)},%
  pdfauthor={D. Kikola et al.},%
  pdfsubject={},%
  pdfkeywords={Quarkonium Production, Spin, TMDs Fixed-Target experiment, Large Hadron Collider},%
  pdfstartview={},%
  bookmarksopen=true, breaklinks=true, debug=true, %
  colorlinks=true, linkcolor=red, citecolor=blue, urlcolor=blue
}

\journalname{Few Body Systems}
\begin{document}

\title{Feasibility Studies for Single Transverse-Spin Asymmetry Measurements at a Fixed-Target Experiment Using the LHC Proton and
Lead Beams (AFTER@LHC)}

\author{D.~Kiko{\l}a \and M.~G.~Echevarria \and C.~Hadjidakis \and J.P.~Lansberg \and C.~Lorc\'e \and L.~Massacrier \and C.~Quintans \and A.~Signori \and B.~Trzeciak}

\institute{  
D. Kiko{\l}a \at Faculty of Physics, Warsaw University of Technology,  Warsaw, Poland \\
M. G. Echevarria \at FQA and ICCUB, Universitat de Barcelona, Mart\'i i Franqu\`es 1, 08028 Barcelona, Spain\\
C. Hadjidakis, J.P. Lansberg, L. Massacrier \at IPNO, CNRS-IN2P3, Univ. Paris-Sud, Universit\'e Paris-Saclay, 91406 Orsay Cedex, France\\
C. Lorc\'e \at CPhT, Ecole Polytechnique, CNRS, Universit\'e Paris-Saclay, Palaiseau, France\\
C. Quintans \at LIP, Lisbon, Portugal\\
A. Signori \at Theory Center, Thomas Jefferson National Accelerator Facility, 12000 Jefferson Avenue, Newport News, VA 23606, USA\\
B. Trzeciak \at Institute for Subatomic Physics, Utrecht University, Utrecht, The Netherlands\\
}

\date{\vspace*{-6.9cm} {\begin{flushright} JLAB-THY-17-2401 \end{flushright}\vspace*{7.cm}}}

\maketitle

\begin{abstract}
	
The measurement of Single Transverse-Spin Asymmetries, $A_N$, for various quarkonium states and Drell-Yan lepton pairs can shed light on the orbital angular momentum of quarks and gluons, a fundamental ingredient of the proton-spin puzzle. 
The \AFTER\ proposal combines a unique kinematic coverage and large luminosities thanks to the Large Hadron Collider beams to deliver precise measurements, complementary to the knowledge provided by collider experiments such as at RHIC. In this paper, we report on sensitivity studies for \jpsi, \ups\ and Drell-Yan $A_N$ done using the performance of LHCb-like or ALICE-like detectors, combined with polarised gaseous hydrogen and helium-3 targets. 
In particular, such analyses will provide us with new insights and knowledge about transverse-momentum-dependent parton distribution functions for quarks and gluons and on twist-3 collinear matrix elements in the proton and the neutron. 
 
\end{abstract}

\tableofcontents

\section{Introduction}
\label{intro}

\subsection{Single Transverse-Spin Asymmetries: what for?}

The spin is a fundamental quantity of a nucleon, yet its origins are largely unknown. It has been a topic of intense theoretical and experimental studies since the European Muon Collaboration reported that the spin of constituent quarks only accounts for a small fraction of the observed spin, $1/2$~\cite{Ashman:1987hv}. Nowadays, it is accepted that both quarks and gluons as well as their relative motion, via the orbital angular momentum (OAM), contribute to the nucleon spin. For instance, in the case of a longitudinally polarised nucleon, \ie\ with helicity $+\frac{1}{2}$, the spin is given by a sum rule
\begin{equation}
\frac{1}{2} = \frac{1}{2}\Delta \Sigma + \Delta G + {\cal L}_{q} + {\cal L}_{g}
\,,
\label{e:spindec}
\end{equation}
where $\frac{1}{2}\Delta\Sigma$ is the combined spin contribution of the quarks and the antiquarks, $\Delta G$ is the gluon spin, and ${\cal L}_{q,g}$ are the quark and gluon OAM contributions. Recent experimental data have shown that the spin distributions of quarks and antiquarks only account for about 25\% of proton total longitudinal spin~\cite{deFlorian:2008mr}, and that of the gluons for about 20\% for $x>0.05$~\cite{Adamczyk:2014ozi}, yet compatible with zero. 
The remaining proton spin therefore should arise from the relative dynamics of quarks and gluons, \ie\ via ${\cal L}_{q}$ and ${\cal L}_{g}$. 
Understanding the parton transverse dynamics should then shed light on the origin of the proton spin.

The transverse-spin distributions give access to the aforementioned intrinsic properties 
of the proton constituents: their transverse and orbital-angular momenta. 
Contrary to the quark sector, very little is known about the gluon contributions via
$\Delta G$ and ${\cal L}_{g}$ to the transverse spin. Only recently, COMPASS 
observed a non-zero asymmetry on the order of $20\%$ ($2\sigma$ away from zero)
which hints at a non-zero value of ${\cal L}_{g}$~\cite{Kurek:2016nqt,Adolph:2017pgv}.
Indeed, whereas Single Transverse-Spin Asymmetries (STSAs), \An, are not directly connected to ${\cal L}_{q,g}$, a non-zero \An\
imposes that ${\cal L}_{q,g}$ is non-zero.

Such studies require gluon-sensitive observables. Naturally, quarkonium production can serve as such 
tool, since the gluon fusion is the dominant contribution to these processes in high-energy hadron 
collisions (see~\cite{Feng:2015cba,Brodsky:2009cf,Lansberg:2010vq} for RHIC energies). Thanks to many different experiments performed in the 
last decades (for reviews see~\cite{Andronic:2015wma,Brambilla:2010cs,Lansberg:2006dh}), 
measuring quarkonia via leptonic decays became a relatively 
straightforward task. The downside remains a lower production cross-section 
compared to light mesons, which calls for large integrated luminosities. 
Lastly, accessing transverse-spin physics research requires a polarised target.

At the moment, only the Relativistic Heavy Ion Collider (RHIC) at the Brookhaven 
National Laboratory provides polarised \pp\ collisions at large enough energies 
for STSA measurements for quarkonia. Indeed, 
the PHENIX collabration~\cite{Adare:2010bd} pioneered the studies of $p^\uparrow p \to J/\psi X$. 
The precision of the measurement is however very limited and did not allow to 
claim for a non-zero STSA, or to constrain ${\cal L}_{g}$. 
The COMPASS detector is probably the other existing experimental set-up 
where quarkonium STSA can be measured.
Measurements of other quarkonium states are extremely important. First, 
$J/\psi$ production involves a complicated pattern of feed downs and 
vector quarkonium production is not the easiest case to analyse theoretically.
C-even quarkonia  $(\chi_{c,b}, \eta_c)$ which can be produced alone 
without recoiling gluons indeed offer some advantages. We refer 
to~\cite{Boer:2016bfj,Schafer:2013wca,Boer:2012bt,Yuan:2008vn} for more details.

\subsection{\AFTER : why ?}

\AFTER\ is a proposal~\cite{Brodsky:2012vg} for a fixed-target experiment using 
the LHC proton and heavy-ion beams at high and continuous luminosities, yet 
parasitic to the other LHC experiments. It bears on a particularly wide range of 
physics opportunities~\cite{Koshkarev:2016ket,Signori:2016lvd,Koshkarev:2016acq,Lansberg:2016gwm,Lansberg:2016urh,Signori:2016jwo,Pisano:2015fzm,Vogt:2015dva,Fengand:2015rka,Barschel:2015mka,Kikola:2015lka,Kurepinand:2015jka,Zhou:2015wea,Arleo:2015lja,Lansberg:2015lva,Brodsky:2015fna,Massacrier:2015qba,Anselmino:2015eoa,Lansberg:2015kha,Ceccopieri:2015rha,Goncalves:2015hra,Kanazawa:2015fia,Lansberg:2015hla,Massacrier:2015nsm,Lansberg:2014myg,Chen:2014hqa,Rakotozafindrabe:2013au,Lansberg:2013wpx,Lansberg:2012sq,Lorce:2012rn,Rakotozafindrabe:2012ei,Boer:2012bt,Lansberg:2012wj,Lansberg:2012kf,Liu:2012vn} thanks to the typical boost of the fixed-target mode between the center-of-mass (c.m.s.) and the laboratory frame converting
forward detectors into backward ones, 
to relatively high luminosities and not too low c.m.s. energies, 115 GeV per nucleon with the 7 TeV proton beam and 72 GeV per nucleon with the 2.76 TeV lead beam. 
In this energy range, the cross section for quarkonium production is already high and gluon fusion dominates. 
Larger charmonium, bottomonium, $D$, $B$ and Drell-Yan pair yields are expected compared to previous
fixed-target experiments.

\AFTER\ will also give access to the target-fragmentation region $x_F \to -1$, with detectors similar to the ALICE or LHCb ones, enabling the exploration of large momentum fractions in the target. 
Moreover, it is relatively easy to use a polarised gas target~\cite{Steffens:2015kvp} 
in one of the existing LHC experiments keeping high integrated luminosities.
Doing so, spin studies in the gluon sector via gluon-sensitive probes become more than possible
with the investigation of spin correlations such as the Sivers effect~\cite{Sivers:1989cc,Brodsky:2002rv,D'Alesio:2007jt,Barone:2010zz} or the correlation between the gluon transverse momentum (denoted $k_T$ thereafter) and the nucleon spin.

All this allows for measurements of gluon sensitive probes with an unprecedented quality
in a region, large $x^\uparrow$, where theory calculations~\cite{Anselmino:2015eoa} predict that the effect (the 
observed spin-correlated azimuthal modulation of the produced particles) is the largest.
We stress that the same observation obviously holds for open heavy flavour and Drell-Yan pairs.

The structure of the paper is as follows. In the next section, we recall some
theoretical concepts related to spin studies with \AFTER. Next, we present 
feasibility studies for STSAs and prospects for other spin-related measurements. 
The last section gathers our conclusions.

\section{Theory}
\label{theory}
In order to measure the parton OAM, one should consider observables which are sensitive to both the parton transverse position and momentum. 
These are usually related to the Generalised Parton Distributions, accessible via exclusive processes. 
However, one can also indirectly obtain information on the orbital motion of partons  via STSAs in hard-scattering processes, where one of the colliding hadrons is transversely polarised (see e.g.~\cite{D'Alesio:2007jt,Barone:2010zz} for recent reviews). 
These asymmetries are naturally connected to the transverse motion of partons inside hadrons~\cite{Angeles-Martinez:2015sea}. 

The STSA, denoted by $\An$, is the amplitude of the spin-correlated azimuthal modulation of the produced particles:
\begin{equation}
A_{N} = \frac{1}{P}\frac{\sigma^{\uparrow} - \sigma^{\downarrow}}{\sigma^{\uparrow} + \sigma^{\downarrow}}
\,,
\end{equation}
where $\sigma^{\uparrow\,(\downarrow)}$ is the differential cross section (or 
yield) of particles produced with the target spin polarised upwards (downwards) 
with respect to the incoming beam direction, and $P$ is the effective target polarisation.
Studying $\An$ is of particular interest because leading-twist collinear perturbative QCD 
predicted it to be small ($\An \propto m_q/p_T \sim O(10^{-4})$), while the 
measured $\An$ was observed to be $\sim 10\%$ or even larger at high $x_F$ 
in polarised collisions over a broad range of energies~\cite{Bonner:1988rv,
Adams:1991rw,Adare:2013ekj}. For example, at Fermilab STSAs on the order of $10\%$ 
were measured in hadronic polarised hyperon production~\cite{Lesnik:1975my,Bunce:1976yb}
 and pion production~\cite{Adams:1991rw,Adams:1991cs,Adams:1991ru,Krueger:1998hz}. 

Since then, the study of spin asymmetries has rapidly evolved, both from the 
theoretical and experimental point of view (for detailed insights into spin 
physics and azimuthal asymmetries, see e.g. \cite{Barone:2003fy,Leader:2001gr,
Pijlman:2006vm,Bomhof:2007zz,Barone:2001sp,Bourrely:1980mr,Anselmino:1994gn,
Liang:2000gz,Boglione:2015zyc,Aschenauer:2015ndk,Prokudin:2014pea}). 
This triggered investigations of the hadron structure beyond the 
collinear parton model, and different mechanisms were proposed to account 
for spin asymmetries~\cite{Pijlman:2006vm,Bomhof:2007zz}.

The first interpretation of 
STSAs relied on a collinear factorisation framework~\cite{Efremov:1981sh,Qiu:1991pp}, 
involving interactions of gluons from the target remnants with the active partons 
in the initial and final states. This is accounted for by collinear twist-$3$ (CT3) 
matrix elements, the so-called Efremov-Teryaev-Qiu-Sterman (ETQS) matrix elements. Later, Sivers 
proposed an explanation~\cite{Sivers:1989cc,Sivers:1990fh} based on a correlation 
between the transverse momentum of the quark and the polarisation of the proton, 
introducing the quark transverse momentum dependent parton distribution function 
(TMD PDF) $f_{1T}^\perp(x,k_T^ 2)$ (the so-called Sivers function). The common 
feature of these two mechanisms is that an imaginary phase required for the 
non-vanishing asymmetry is generated by taking into account an additional gluon 
exchange between the active parton and the remnant of the transversely polarised 
hadron. 

The CT3 formalism is valid for processes with only one characteristic hard scale, 
for instance the transverse momentum of a 
produced hadron, satisfying $p_{hT} \gg \Lambda_{\rm QCD}$, in a proton-proton collision. 
The TMD formalism~\cite{Collins:2011zzd,
GarciaEchevarria:2011rb,Echevarria:2012js,Echevarria:2014rua,Echevarria:2015uaa},
on the other hand, is suited for processes with two characteristic and 
well-separated scales (for example, in Drell-Yan process, the mass $M$ and the 
transverse momentum $p_{T}$ of the produced lepton pair, where 
$\Lambda_{\rm QCD} \lesssim p_{T} \ll M$)~\footnote{See also \eg~\cite{DAlesio:2004eso,Anselmino:2004ky} for what is known 
in the literature as the generalised parton model, which is an extension of the collinear 
perturbative QCD approach to incorporate the transverse dynamics of partons, and 
resembles the TMD formalism from a more phenomenological perspective.}
. 
When the two relevant scales become comparable, the TMD formalism can be then reduced to the CT3 one.
In practice, this is realised in terms of an operator product expansion, since 
the Sivers TMD function can be matched onto the ETQS matrix elements at large 
transverse momenta. Thus, depending on the process, $\An$ should be addressed 
either using the CT3 formalism through 3-parton correlation functions, or the 
TMD formalism through the Sivers function.

One of the most important predictions, shared by both approaches, is the 
predictable, but non-universal, magnitude of this asymmetry in different process.
The experimental check of this feature is one of the milestones of the $\AFTER$ 
spin physics program.

%%%%%%%%%%%%%%%%%%%%%%%%%%%%%%%%%%%%%%%%%%%%%%%%%%%%
\subsection{Quark Sivers effect}
\label{ss:quark_sivers}
%%%%%%%%%%%%%%%%%%%%%%%%%%%%%%%%%%%%%%%%%%%%%%%%%%%%

Drell-Yan (DY) lepton-pair production is a unique playground to understand the 
physics underlying the Sivers effect: it is theoretically very well understood  
and the quark Sivers function $f_{1T}^{\perp q}(x,k_T^2)$ enters the differential cross sections for DY and semi-inclusive deep inelastic scattering (SIDIS) with opposite sign~\cite{Boer:2003cm}:
\begin{equation}
f_{1T}^{\perp q}(x,k_T^2)\Big|_{\rm DY} = - f_{1T}^{\perp q}(x,k_T^2)\Big|_{\rm SIDIS}
\,.
\end{equation}
$f_{1T}^{\perp q}(x,k_T^2)$ accounts for the number density of unpolarised quarks carrying a longitudinal fraction $x$ of the proton momentum and with transverse momentum $\kT$ for a given transverse spin of the proton $S_T$. 
The verification of this "sign change" is the main physics case of the DY COMPASS run~\cite{Quintans:2011zz} and the experiments E1039~\cite{Klein:zoa} and E1027~\cite{Isenhower:2012vh} at Fermilab.
$\AFTER$ is a complementary facility to further investigate the quark Sivers effect by measuring DY STSAs~\cite{Liu:2012vn,Anselmino:2015eoa} over a wide range of $x^\uparrow$ in a single set-up.
With the high precision that $\AFTER$ will be able to achieve, it will clearly consolidate previous possible measurements.
In the case the asymmetry turns out to be small and these experiments cannot get to a clear answer, 
$\AFTER$ could still confirm or falsify this sign-change prediction and put strict constraints on the Sivers effect for quarks.

In addition, given that this asymmetry can be framed as well within the CT3 approach when the transverse momentum of the produced lepton pair is comparable to its mass, $\AFTER$ will also generate very useful data to constrain the ETQS 3-parton correlation functions.
The latter can also be determined by using direct $\gamma$ production~\cite{Qiu:1991wg}.

%%%%%%%%%%%%%%%%%%%%%%%%%%%%%%%%%%%%%%%%%%%%%%%%%%%%
\subsection{Gluon Sivers effect}
\label{ss:gluon_sivers}
%%%%%%%%%%%%%%%%%%%%%%%%%%%%%%%%%%%%%%%%%%%%%%%%%%%%

The gluon Sivers function is more involved than its quark analogue: different 
processes probe different gluon Sivers functions, due to the inherent process 
dependence of this TMD function~\cite{Buffing:2013kca,Boer:2015vso}. However all 
of them can be expressed in terms of only two independent 
functions~\cite{Pisano:2013cya,Buffing:2013kca}, which will appear in different 
combinations depending on the process. $\AFTER$ will prove to be extraordinary 
useful in disentangling them and testing this generalised universality.

Drell-Yan lepton-pair production is the golden process to access the intrinsic 
transverse motion of quarks in a nucleon. However, there is no analogous process, 
which is at the same time experimentally clean and theoretically well-controlled, 
to probe the gluon content. One of the best tools at our disposal is the 
production of quarkonium states and open heavy-flavour mesons, a major strength 
of $\AFTER$. They provide final states with a typical invariant mass ($M_{\cal Q}$) 
which is, on the one hand, small enough to be sensitive to the intrinsic transverse 
momenta of gluons ($k_T$), and on the other, large enough to realise the hierarchy 
of scales ($M_{\cal Q}\gg k_T$), and thus to allow for the TMD formalism to be applied 
without pollution from higher-twist effects. To this extent, production of C-even 
states can be fruitfully investigated~\cite{Boer:2014tka,Boer:2012bt,
Echevarria:2015uaa,Signori:2016jwo,Signori:2016lvd,Schafer:2013wca}. 

The hadroproduction of $\eta_c$ has already been measured by LHCb at high transverse momentum above 
$p_T=6$~GeV~\cite{Aaij:2014bga}, as well as non-prompt $\eta_c(2S)$~\cite{Aaij:2016kxn}.
With an LHCb-like detector, STSAs for $\chi_c$, $\chi_b$ and $\eta_c$ are at reach,
 as demonstrated by studies of various $\chi_c$ states~\cite{LHCb:2012af,LHCb:2012ac}
 in the busier collider environment down to $p_T$ as low as 2 GeV. 
At lower energies, the reduced combinatorial background will give access to lower $p_T$. 
Moreover, with $\AFTER$, the production of $\jpsi$, $\psip$ and $\ups$ will also 
allow for accurate measurements of the gluon Sivers effect, as it is shown in 
the next Section.

The open heavy-flavour production also allows to investigate the process dependence
 of $\An$ (measuring charm quarks and anti-quarks separately)~\cite{Kang:2008ih}. 
Moreover, it is a unique probe of C-parity odd twist-3 tri-gluon 
correlators~\cite{Ji:1992eu,Beppu:2010qn}, for which $\AFTER$ will obtain valuable
 information.

Finally, momentum imbalance observables also provide a very useful handle to probe
 the gluon Sivers function and its $k_T$ dependnce. $\jpsi + \gamma$ production,
 is probably one of the cleanest from the theoretical point of view~\cite{Dunnen:2014eta}
along with di-$J/\psi$ production.

\section{Feasibility studies for quarkonium and DY \An}
\label{subsec:SSA}
\subsection{Simulation setup}

We consider here two possible options for the realisation of \AFTER\ with a polarised target: using the LHCb detector~\cite{Alves:2008zz} or the ALICE detector~\cite{Aamodt:2008zz}. LHCb is a multi-purpose, single arm detector with a precise microvertexing instrumentation, particle identification systems, electromagnetic and hadronic calorimeters. 
It has capabilities for an accurate primary and secondary vertex location determination, for a particle identification (including $\pi, K, p, e$ and $\mu$) with a good momentum resolution and a high rate data acquisition system. LHCb has already successfully run in the fixed-target mode using its luminosity monitor SMOG (System for Measuring the Overlap with Gas)~\cite{Aaij:2014ida} as a gas target. However, the data taking was done
over limited periods of time and limited gas pressures. Additional feasibility studies are needed to address the possibility of installing a polarised gas target. The simulation setup we will use for a LHCb-like detector is described in detail in~\cite{Massacrier:2015qba}. We therefore only outline here the most relevant information for Drell-Yan and $\jpsi\,,\ups \rightarrow \mu^+ \mu^-$ measurements, namely: the single muon tracking and the identification efficiency is $\sim 98\%$ and the kinematic acceptance is $2<\eta < 5$ and a $\pT$ threshold is set at $\pT > 0.7 \, \gevc$. However, for the studies presented in this paper, we used a more stringent cut of $\pT > 1.2 \, \gevc$ which usefully reduces the uncorrelated background.

For ALICE, a target can in principle be placed, either at the nominal interaction point ($IP_Z = 0$), or upstream from the nominal one (for instance at 5~m, $IP_Z = -5$~m). The Muon Spectrometer would then provide a setup for di-muon pair measurements with a single track acceptance of $2.5<\eta < 4$ for $IP_Z = 0$, and $3.2<\eta < 4.2$ for $IP_Z = -5$~m. The Muon Forward Tracker (MFT)~\cite{CERN-LHCC-2015-001}, which will be installed in the near future, will add tracking capabilities for muon measurements for $IP_Z = 0$. A typical single track $\pT$ threshold for di-muon pair analysis with the ALICE Muon Arm is $\pT > 1 \, \gevc$~\cite{Adam:2015rta}. In addition, the muon arm is equipped with an absorber, which reduces the uncorrelated background due to misidentified hadrons in such studies. Moreover, it is possible to install an additional vertexing instrumentation near the interaction point for $IP_Z = -5$~m. Such a vertex detector would improve the precision of Drell-Yan measurements by removing the background muons from light-hadron and charm/bottom-hadron decays. Indeed, for $IP_Z = -5$~m, the acceptance of the MFT does not match that
of the spectrometer.

To quantify the STSA, we use the amplitude of the spin-correlated azimuthal modulation of the produced particles $\An$.
We consider the following approach to the $\An$ measurement with di-muon pairs. First, a microvertexing detector will allow to remove the correlated background from charm and bottom hadron decays. Thanks to the boost effect, their decay vertex is well separated from the primary collision point and $c \rightarrow \mu$ and $b \rightarrow \mu$ will be identified and then removed from the pairs used to construct the invariant mass distribution. 
For the statistical precision evaluation, we assume that the yields $\sigma^{\uparrow}, \sigma^{\downarrow}$ are measured separately with a standard invariant mass approach used in high energy experiments for quarkonium studies. We construct an invariant mass spectrum of all di-muon pairs (often called \textit{foreground}) which contains both signal of interest (\jpsi, \ups, DY) and random pairs (combinatorial background, B). The background yield B is evaluated with a like sign technique: by taking a sum of mass distributions for $\mu^+\mu^+$ and $\mu^-\mu^-$,  or a geometric mean of like-sign pair yields ($2\sqrt{ N_{\mu^+\mu^+} N_{\mu^-\mu^-} }$). Finally, we subtract the background from the foreground to get the signal yield for a given target polarisation in a given kinematical domain. PHENIX used such an approach in \jpsi\ $\An$ measurement~\cite{Adare:2010bd} and their study confirmed that the like-sign pairs represent well the yield and polarisation of the uncorrelated background. The statistical uncertainty $\delta_{\sigma}$ on the $\sigma^{\uparrow}, \sigma^{\downarrow}$ is thus given by $\delta_{\sigma} = \sqrt{\sigma + 2B}$, and the statistical uncertainty on $\An$ reads $\delta_{\An} = \frac{2}{P({\sigma^{\downarrow} + \sigma^{\uparrow}})^2} \sqrt{(\delta_{\sigma^{\uparrow}}\sigma^{\downarrow})^2 + (\delta_{\sigma^{\downarrow}}\sigma^{\uparrow})^2}$. The factor $2B$ in $\delta_{\sigma}$ definition accounts for the statistical uncertainty from the combinatorial background subtraction in the $\An$ evaluation. This approach assumes that the luminosities for each polarisation configuration are the same; if not, then the $\sigma^{\uparrow}, \sigma^{\downarrow}$ need to be corrected for the relative luminosity differences.
Similarly, we consider here that the systematic effects (like detector acceptance) are the same for $\sigma^{\uparrow}$ and $\sigma^{\downarrow}$, and will cancel out in the ratio. If they are not, this should be accounted for.

\subsection{Yields and kinematical range for the \An\ measurements with a LHCb-like detector}

\begin{figure}[hbt!]
	\centering
		\subfloat[$2 < y_{\mu^+\mu^-}^{\rm Lab} < 3$]{\includegraphics[width=0.33\textwidth]{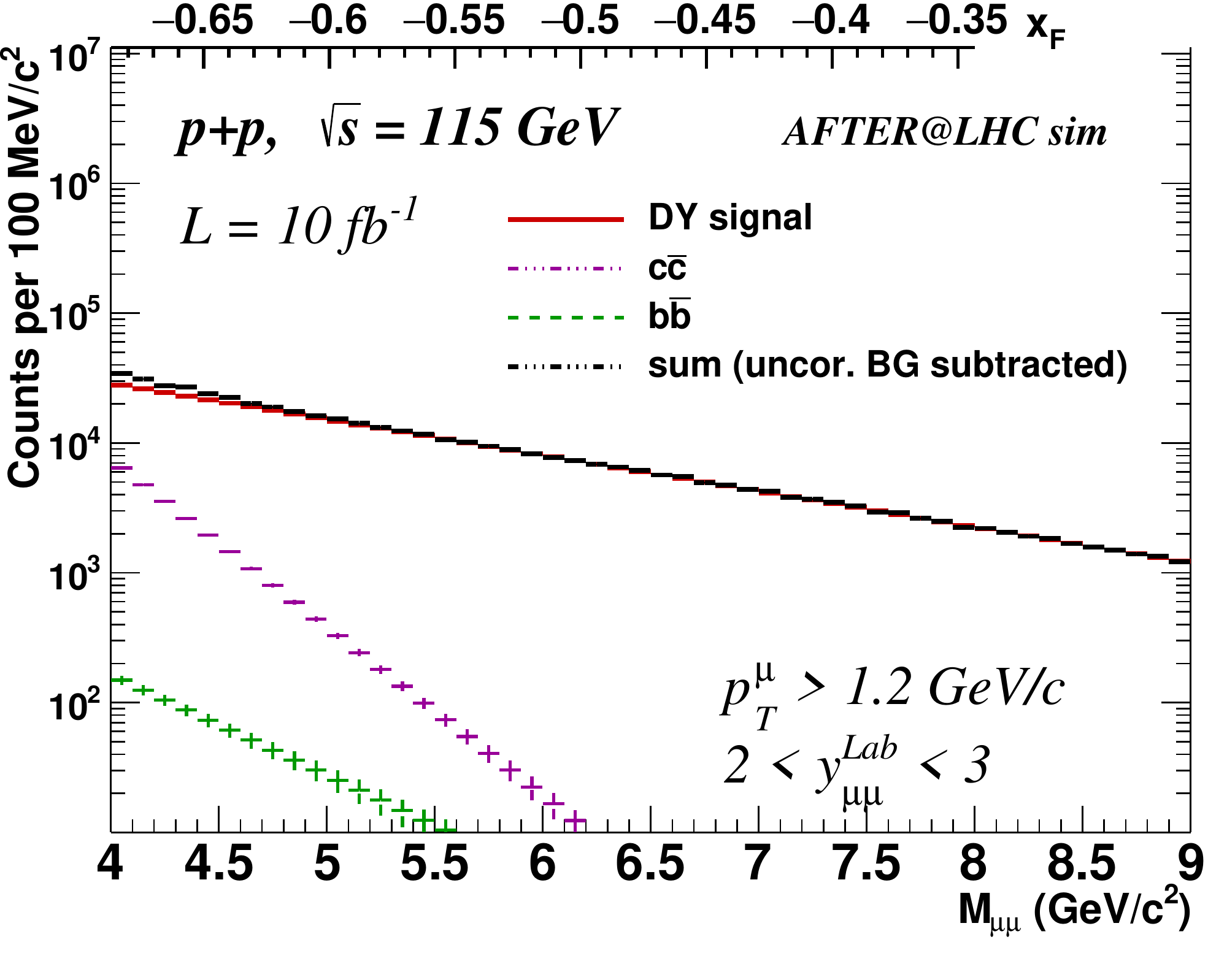}}
		\subfloat[$3 < y_{\mu^+\mu^-}^{\rm Lab} < 4$]{\includegraphics[width=0.33\textwidth]{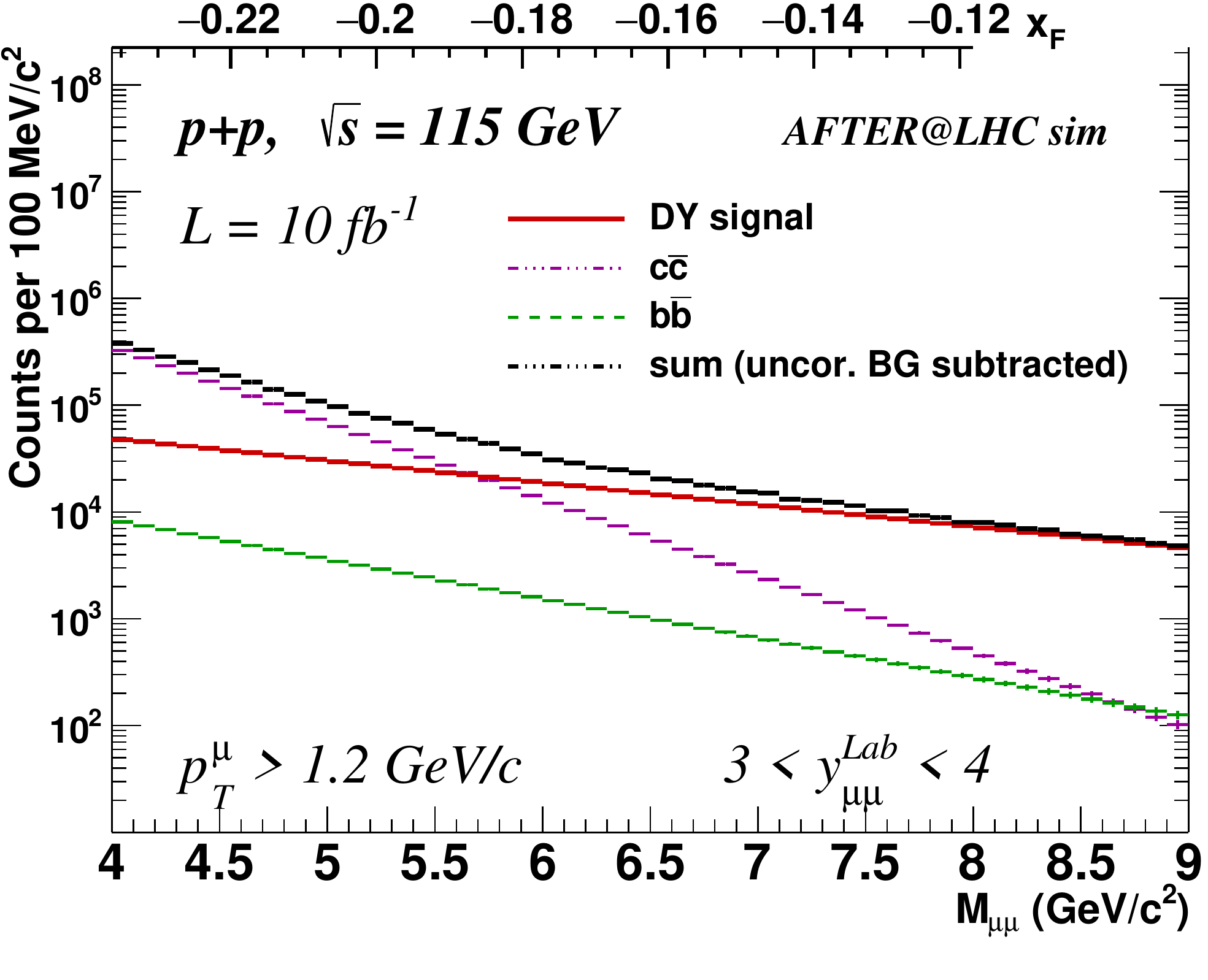}}
		\subfloat[$4 < y_{\mu^+\mu^-}^{\rm Lab} < 5$]{\includegraphics[width=0.33\textwidth]{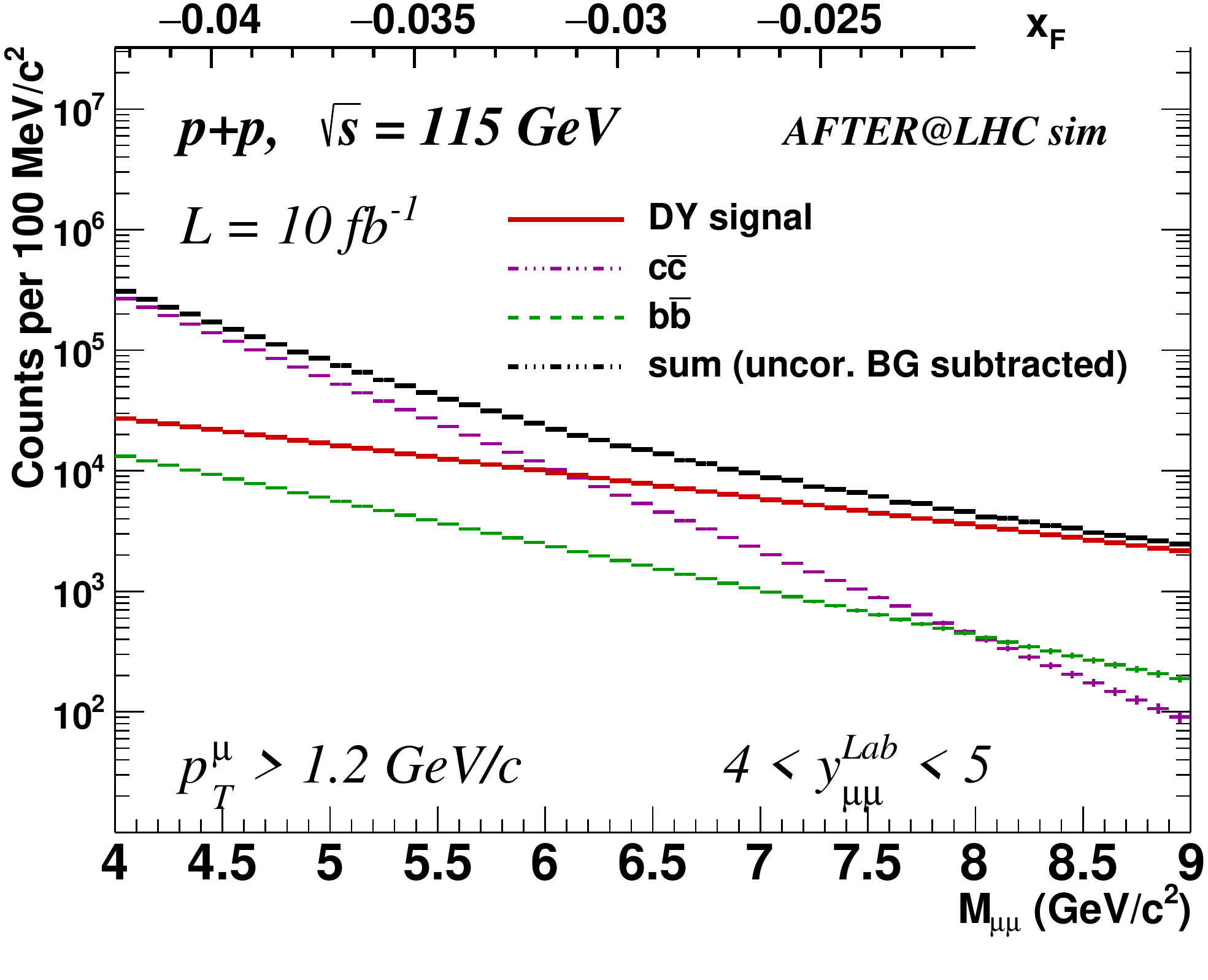}} 
	\caption[dimuon mass disitbution]{Invariant mass distribution of the correlated $\mu^+\mu^-$ pairs originating from Drell-Yan, $c\overline{c}$ and $b\overline{b}$ production for a LHCb-like detector. The uncorrelated background is estimated with the like-sign method and then subtracted from the overall simulated $\mu^+\mu^-$ mass spectrum.}
	\label{fig:DY:mass}
\end{figure}

The statistical precision of quarkonium and Drell-Yan measurements with a LHCb-like detector was obtained with realistic $p+p$ simulations at $\sqrt{s} = 115$ GeV of correlated and uncorrelated background~\cite{Massacrier:2015qba}. First, the invariant mass spectrum of all $\mu^+\mu^-$ pairs ($M_{\rm All}$) is computed, which includes the signal (correlated pairs), the correlated background (mainly muon pairs from semileptonic decays of charmed and bottom hadrons) and the so-called combinatorial background (uncorrelated, randomly combined muon pairs in the analysis). We applied a single-muon $\pT$ threshold of $\pT> 1.2 \, \gevc$ to reduce the background. We estimated the invariant-mass distribution of the uncorrelated background with the like-sign technique and then subtracted it from $M_{\rm All}$ to account for the statistical uncertainty owing to the background fluctuations. Figure~\ref{fig:DY:mass} shows examples of the resulting invariant-mass distributions of the correlated $\mu^+\mu^-$ pairs for three rapidity bins: $2 < y_{\mu^+\mu^-}^{\rm Lab} < 3, 3 < y_{\mu^+\mu^-}^{\rm Lab} < 4$ and $4 < y_{\mu^+\mu^-}^{\rm Lab} < 5$ for a LHCb-like detector. In addition to the Drell-Yan and quarkonium studies, the double-\jpsi\ production is of interest because it gives access to the  $k_T$ evolution of the gluon Sivers function. The double \jpsi\ rates are calculated within LHCb-like acceptance~\cite{Lansberg:2015lva} assuming a negligible background (which is an acceptable approximation given the background level observed for single $J/\psi$~\cite{Massacrier:2015qba}). Table \ref{tab:DY:Quarkonium:yields} \& \ref {tab:DY:StoB} summarises the expected \jpsi, double \jpsi, \ups and Drell-Yan yields for a single data taking year for ${\cal L}_{\rm int} = 10$~fb$^{-1}$.

\begin{table}[!hbt]
	\centering \setlength{\arrayrulewidth}{.8pt}\renewcommand{\arraystretch}{1.1}
	\begin{tabular}{c|ccc}
		\vspace*{-.15cm} \\
			
					  & $2 < y_{\mu^+\mu^-}^{\rm Lab} < 3$ &  $3 < y_{\mu^+\mu^-}^{\rm Lab} < 4$ & $4 < y_{\mu^+\mu^-}^{\rm Lab} < 5$ \smallskip \\
	
		\hline  \hline  \\
		\jpsi               							& $1.69 \cdot 10^7$  & $1.04 \cdot 10^8$  & $1.01 \cdot 10^8$ \\
		
		\hline \hline \\
		
					 & $3 < y_{\mu^+\mu^-}^{\rm Lab} < 4$ &  $4 < y_{\mu^+\mu^-}^{\rm Lab} < 5$ & $3 < y_{\mu^+\mu^-}^{\rm Lab} < 5$ \\
		\cline{2-4} \\
		$\ups(1S)$          							& $4.85 \cdot 10^4$ 	& $8.85 \cdot 10^4$  	& $1.37 \cdot 10^5$ \\
		$\ups(2S)$          							& $9.57 \cdot 10^3$   & $1.85 \cdot 10^4$  	& $2.81 \cdot 10^4$ \\
		$\ups(3S)$          							& $4.35 \cdot 10^3$   & $8.77 \cdot 10^3$  	& $1.31 \cdot 10^4$ \\

		\hline \hline \\

		 & \multicolumn{3}{c} {$2 < y^{\rm Lab} < 5$} \\
		\cline{2-4} \\
		double \jpsi          							& \multicolumn{3}{c} {780}\\

		\hline \hline
	\end{tabular}
	\caption{\jpsi, \ups\ and double \jpsi\ yields expected with a LHCb-like detector per LHC year with a 7 TeV proton beam on a  proton target assuming  ${\cal L}_{\rm int} = 10$~fb$^{-1}$.}
	\label{tab:DY:Quarkonium:yields}
\end{table}

\begin{table}[!hbt]
	\centering \setlength{\arrayrulewidth}{.8pt}\renewcommand{\arraystretch}{1.1}
	\begin{tabular}{l|ccc}

		\vspace*{-.15cm} \\

		& $2 < y_{\mu^+\mu^-}^{\rm Lab} < 3$ &  $3 < y_{\mu^+\mu^-}^{\rm Lab} < 4$ & $4 < y_{\mu^+\mu^-}^{\rm Lab} < 5$ \smallskip \\
		
		\hline  \hline  \\
	             						
		Drell-Yan yield					& $4.32 \cdot 10^5$  & $9.32 \cdot 10^5$  & $4.98 \cdot 10^5$ \\
		Like-sign pairs yield			& $5.84 \cdot 10^5$  & $1.86 \cdot 10^7$  & $4.53 \cdot 10^6$ \\
		Signal-to-Background ratio		& 0.74		   & 0.05		  & 0.11		\\ 
		\hline \hline

	\end{tabular}
	\caption{Drell-Yan yields for $4 < M_{\mu^+\mu^-} < 9 \, \gevcc$ expected with a LHCb-like detector per LHC year with a 7 TeV proton beam on a  proton target 
assuming  ${\cal L}_{\rm int} = 10$~fb$^{-1}$.}
	\label{tab:DY:StoB}
\end{table}

\begin{figure}[hbt!]
	\centering
	
	\begin{tabular}{cc}
	\includegraphics[width=0.4\textwidth]{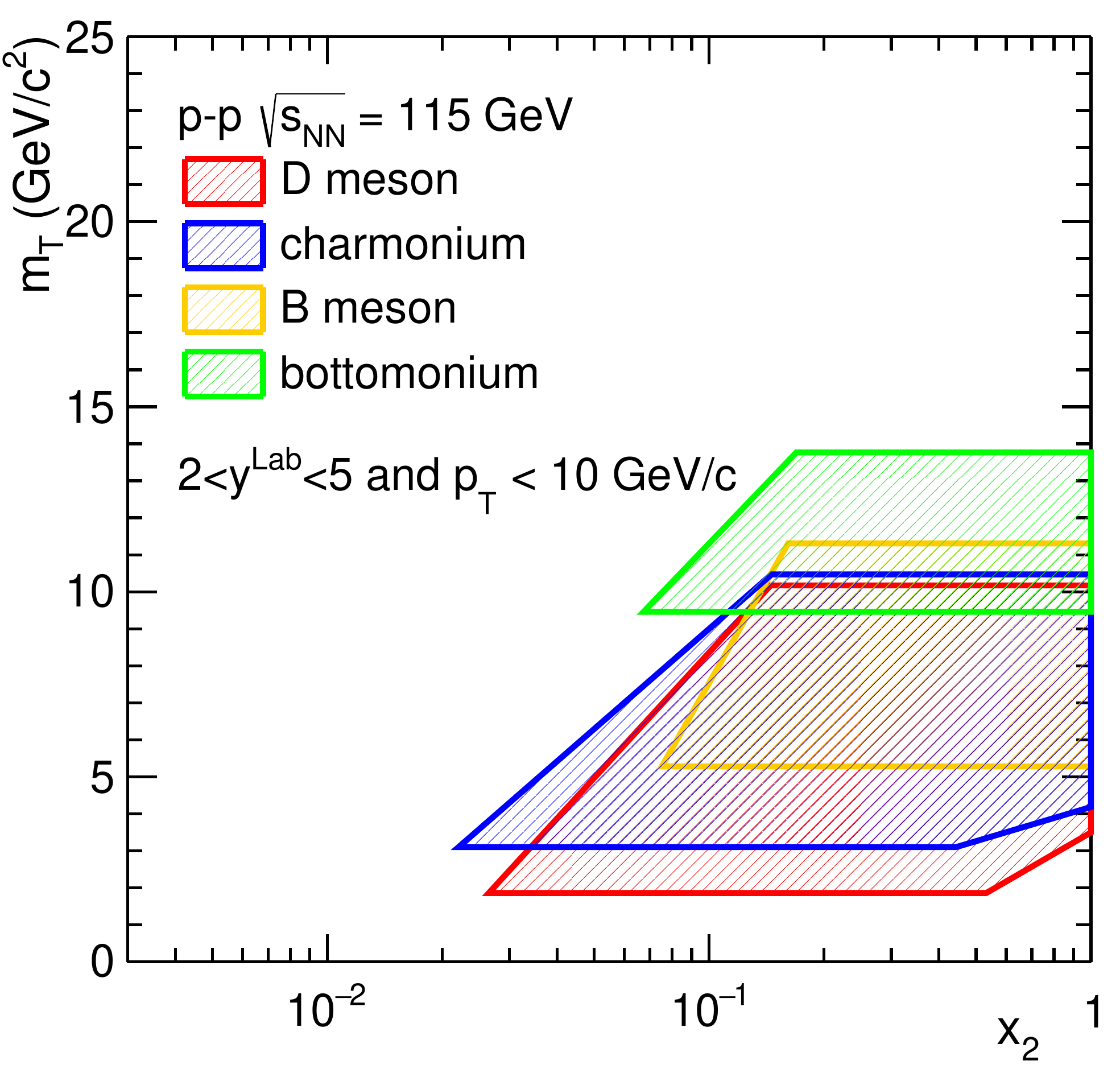} 
	\includegraphics[width=0.6\textwidth]{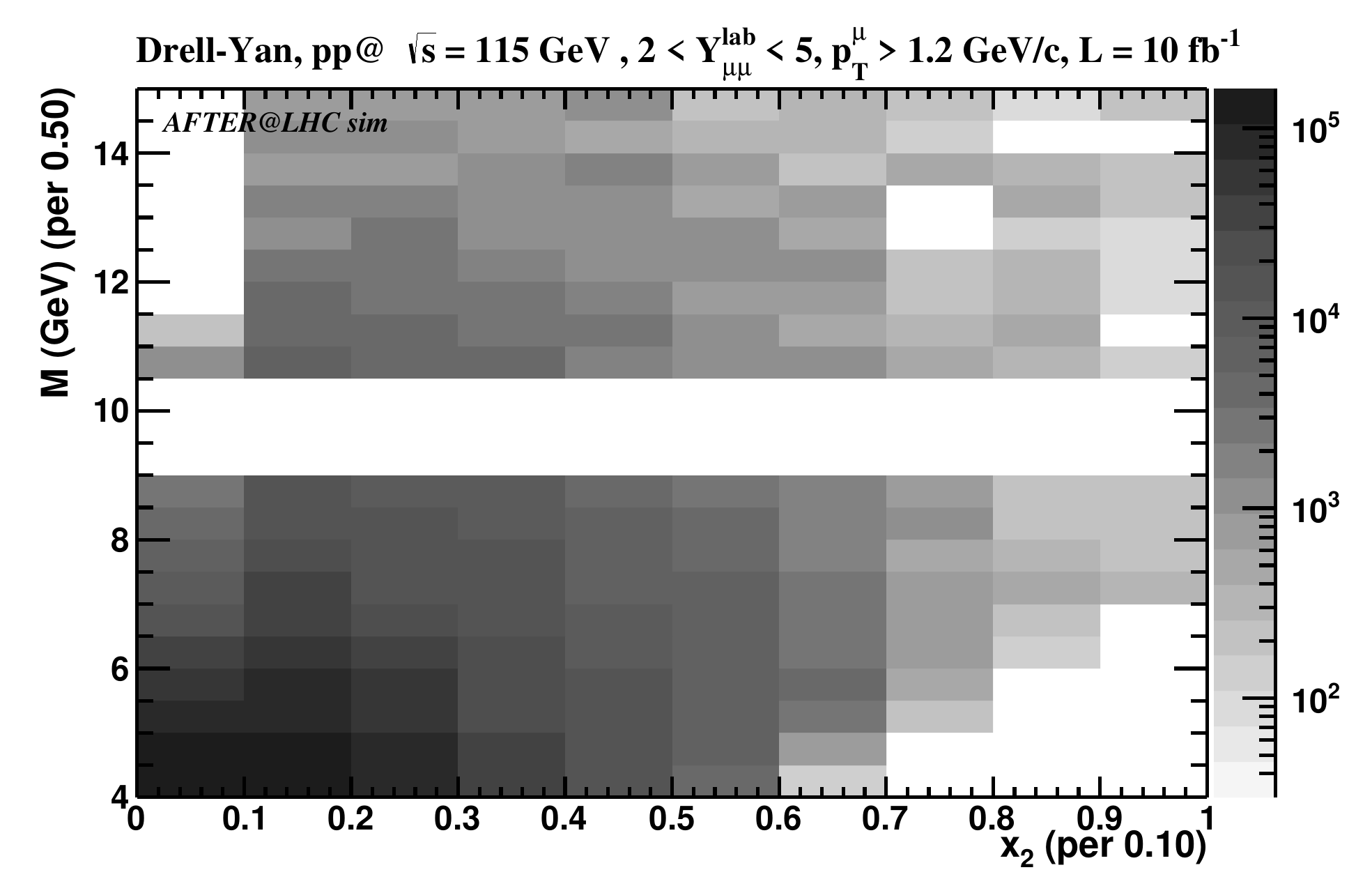} 
		
	\end{tabular}

	\caption[dimuon mass disitbution]{Left panel: The range in the transverse mass $m_{T}$ vs. parton momentum fraction $x_2$ accessible in \AFTER\ in $p+p$ collisions at $\sqrtsNN = 115$~GeV with \jpsi, \ups\ and D-meson measurements. Right panel: the $x_2$ vs. Drell-Yan mass coverage. [Both are for a LHCb-like detector]}
	\label{fig:mass:x2}
\end{figure}

Such an experimental set-up offers a unique kinematic coverage, which allows one to probe a wide range of the momentum fraction $x_2$\footnote{The momentum fractions of the partons $x_1$, $x_2$  are computed with a simplified $2\to 1$ kinematics such 
that   $x_1 = e^y m_T /\sqrt{s}$ and $x_2 = e^{-y} m_T/\sqrt{s}$ with  $m_T^2 = m^2 +p_T^2$}. Figure~\ref{fig:mass:x2} (left panel) shows the range in the transverse mass $m_{T}$ vs. $x_2$ that is available with the \jpsi, \ups\ and B and D-meson measurements. 
Since these are gluon-sensitive probes, they give an unprecedented access to the gluon dynamics over a broad range of $0.02 < x_2 < 1$.  
Figure~\ref{fig:mass:x2} (right panel) shows the corresponding kinematic coverage (mass vs. $x_2$) for the Drell-Yan pairs with the yield information. Any cell in gray contains at least 30 DY events.

\subsection{\An\ in $p+p^{\uparrow}$ collisions.}

Inspired from the performance of the  polarised gas HERMES target~\cite{Steffens:2015kvp} (which successfully operated for many years with an effective average transverse polarisation $P \sim 80\%$), we used $P = 80\%$ in the following statistical precision projections.

\begin{figure}[hbt!]
	\centering
	\begin{tabular}{cc}
		\includegraphics[width=0.48\textwidth]{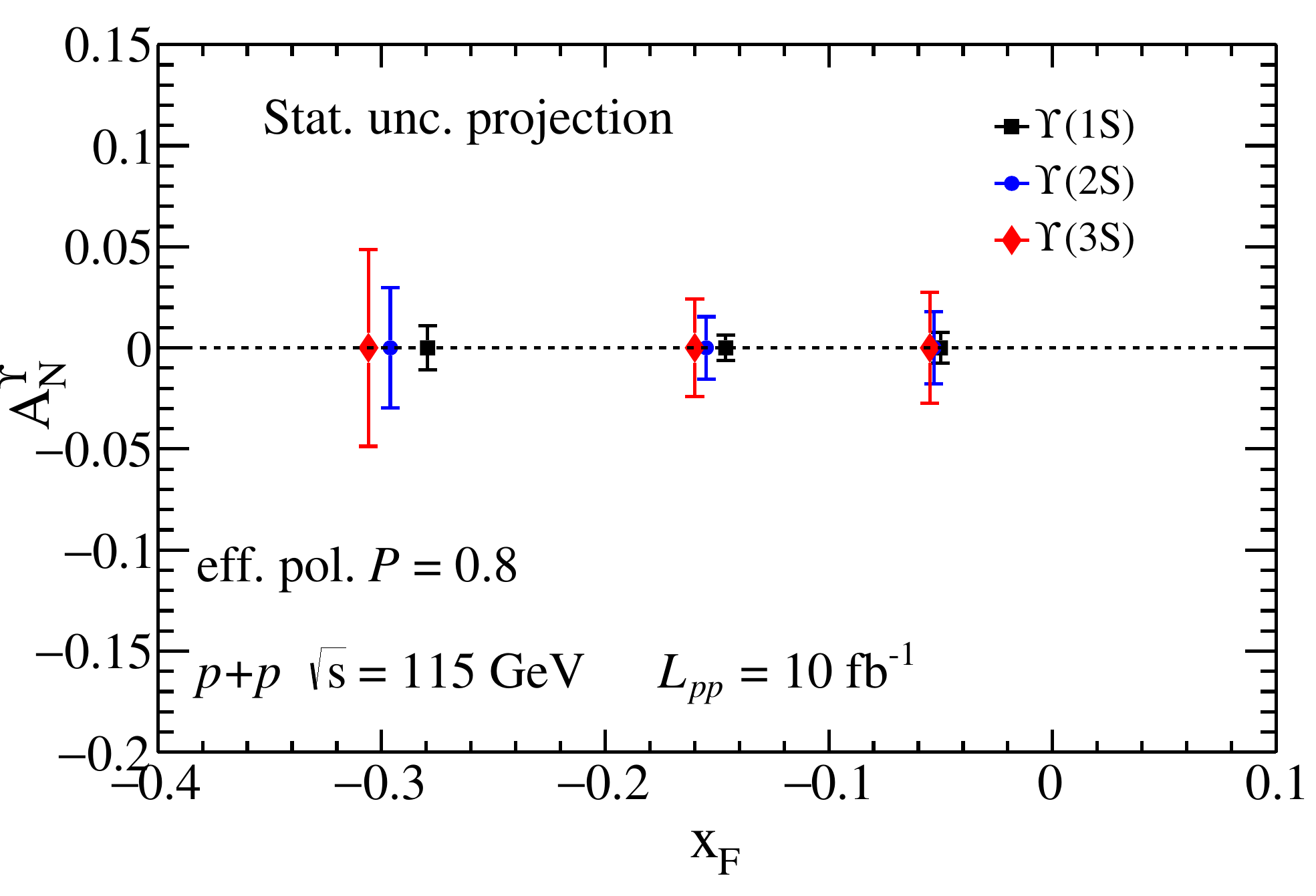} &
		\includegraphics[width=0.48\textwidth]{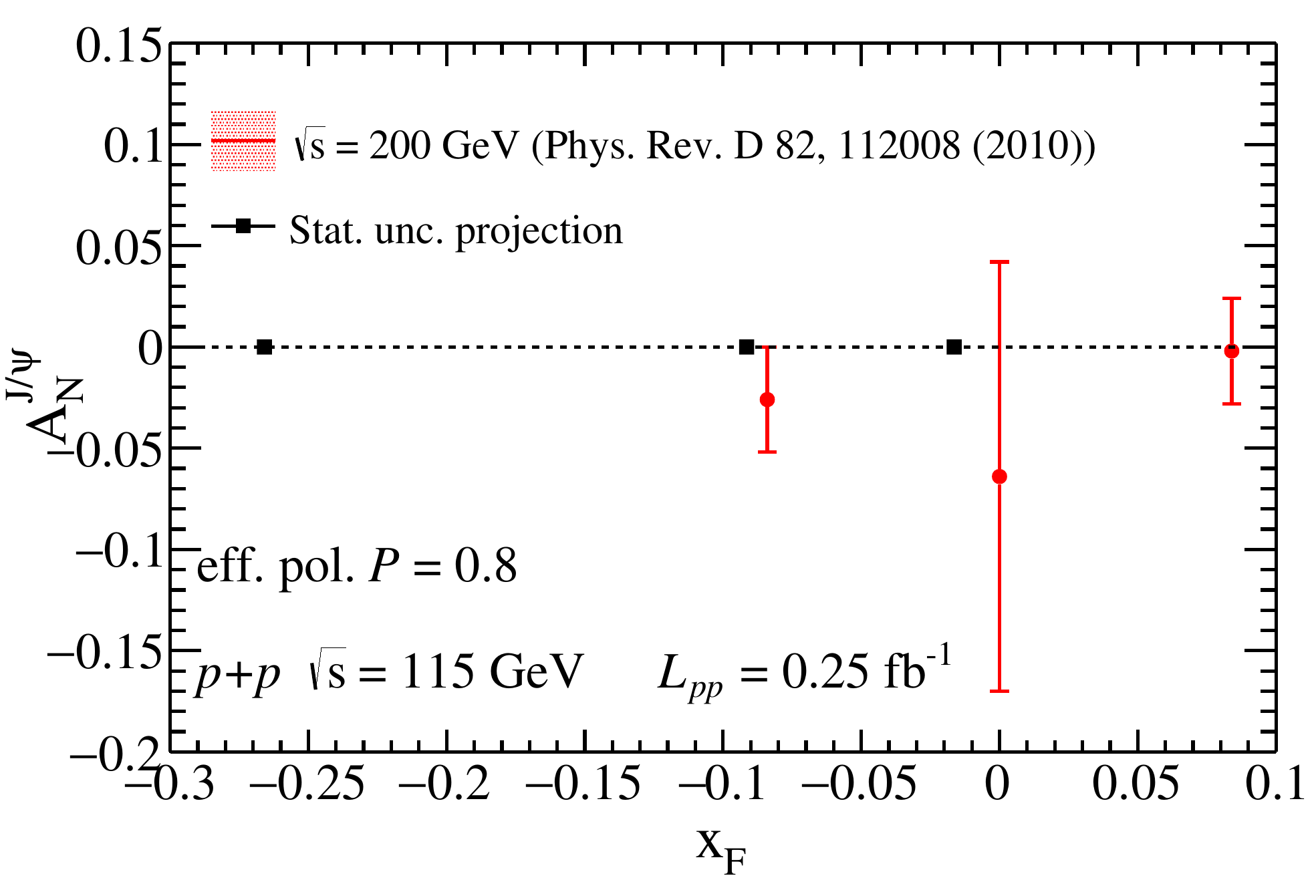} \\
	\end{tabular}
	\caption[Jpsi predictions]{Projections for $\Upsilon$ (left panel) and $\jpsi$ (right panel) $\An$ as a function of $\xF$ for \AFTER\ with a LHCb-like detector. For the $\jpsi$ case,  these are compared to existing data from PHENIX~\cite{Adare:2010bd} in red.}
	\label{fig:An:Jpsi}
\end{figure}

Figure~\ref{fig:An:Jpsi} shows our precision projections for \jpsi\ and \ups\ \An\ as a function of $\xF$. The statistical power of this measurement reflects the major strength of \AFTER, namely an ideal acceptance with conventional detectors and the large production rates for quarkonium states expected for a single year of data taking ($10^6$ $\Upsilon$ and  $10^9$ \jpsi). Such a study will only be limited  by the systematic uncertainties, which cancel out to a large extent in \An. Even for a fraction of the expected luminosity (${\cal L}_{\rm int} = 1$ fb$^{-1}$), $A_N^{J/\psi}$ can be measured with a per-mil precision. Moreover, the $A_N^{\Upsilon(nS)}$ is a unique observable, which is virtually inaccessible elsewhere and which can be measured with a few per cent accuracy with \AFTER. This level of data quality will allow one to study  the size of the asymmetry, its shape and the \xF\ dependence of quarkonium \An. Furthermore, \AFTER\ aims at measurements of \An\ for nearly all quarkonium states, including C-even $\chi_{c,b}$ and $\eta_{c}$, and their associated production. These processes are sensitive to the gluon content of the colliding hadrons which can then be measured with an outstanding precision. 

\begin{figure}[hbt!]
	\centering
	\begin{tabular}{cc}
		\includegraphics[width=0.48\textwidth]{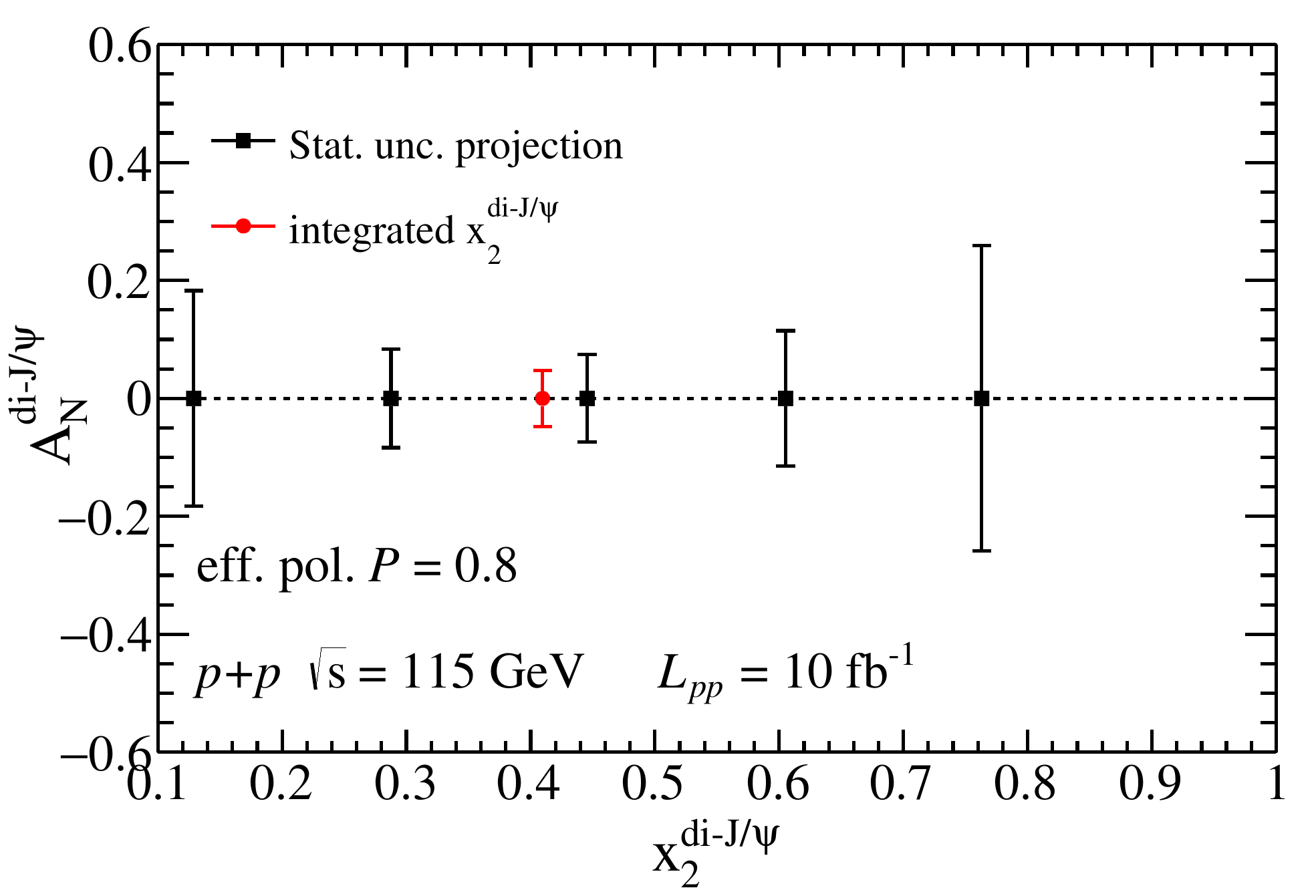} &
		\includegraphics[width=0.48\textwidth]{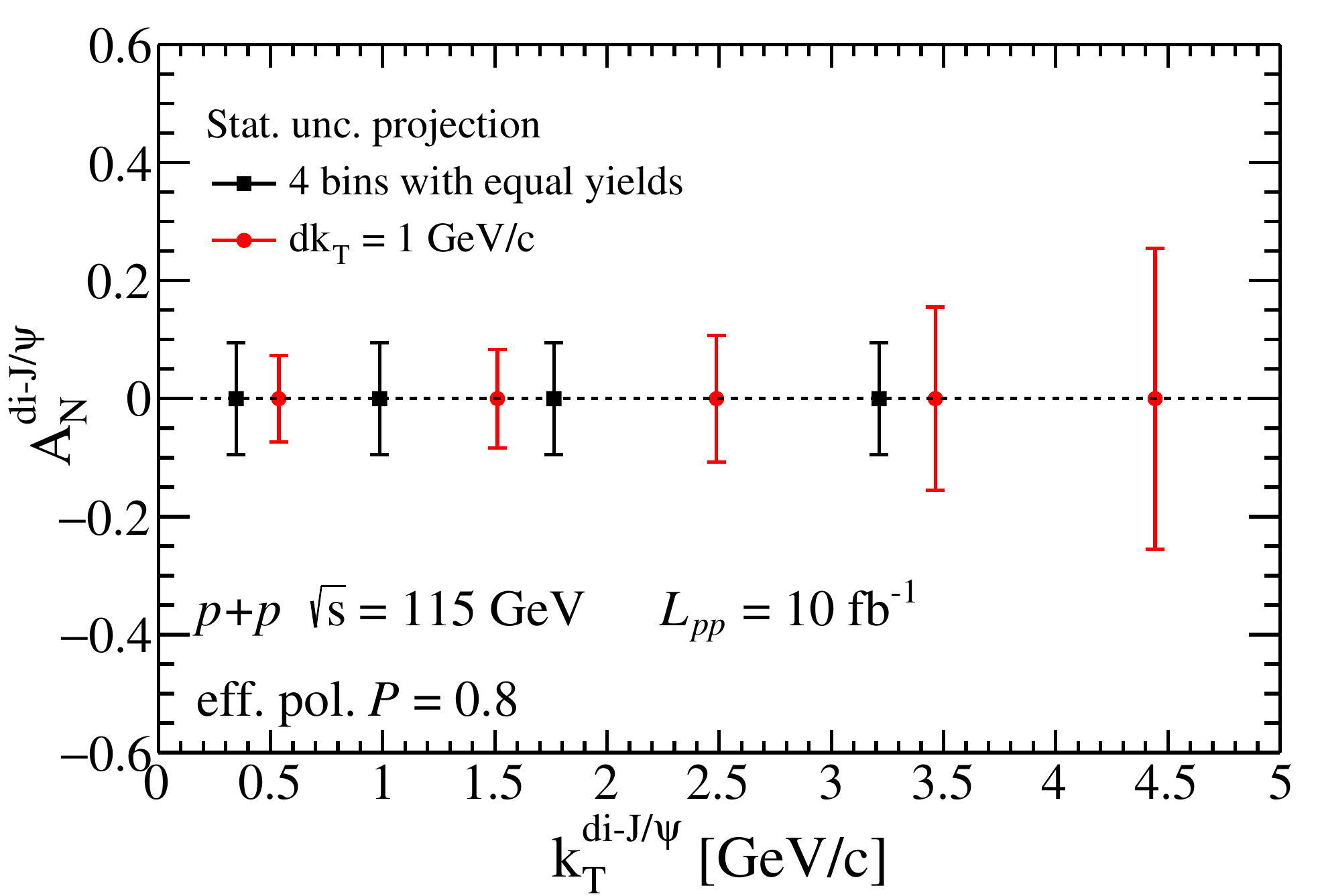} \\
	\end{tabular}
	\caption[Jpsi predictions]{Statistical projections for di-$\jpsi$ $\An$ as a function of $\xF$ and the pair \pT\ with a LHCb-like detector.}
	\label{fig:An:diJpsi}
\end{figure}

Associated-production channels~\cite{Qiu:2011ai,Dunnen:2014eta,Boer:2014lka,Lansberg:2015hla,Signori:2016jwo,Signori:2016lvd,Boer:2016bfj} are fundamental tools to access the Gluon Sivers effect, and also probing the gluon TMD sector and their evolution~\cite{Echevarria:2012pw,Echevarria:2015uaa}.
A few different processes are potentially interesting in this context, for instance $\jpsi-\jpsi$ , $\jpsi-\gamma, \gamma-\gamma$, $\ups-\gamma$. The $\jpsi-\jpsi$ production seems to be the most practical one since the yields are not too small~\cite{Lansberg:2015lva} and the measurement is relatively straightforward (compared, for instance, to direct $\gamma$ studies). 
Figure~\ref{fig:An:diJpsi} shows the \An\ for double \jpsi\ production as a function of the pair $x_2$ and the pair \kT. We consider two scenarios for the analysis of $\An$ as a function of \kT: one with a fixed $k_T$ bin width of 1~\gevc\ ($dk_T = 1\, \gevc$, red points) and four bins with equal yields.
Here, we model the \kT\ dependence as a Gaussian distribution with the width $\sigma = 2 \, \gevc$. 
The $x_2$-integrated \An\ will allow for the determination of the STSA with a few percent precision and the $A_N(k_T)$ gives access --for the first time-- to the $k_T$ dependence of the gluon Sivers TMD up to $\kT \approx 4 \, \gevc$. 

\begin{table}[htb!]
\begin{center} \setlength{\arrayrulewidth}{.8pt}  \renewcommand{\arraystretch}{1.5}
%\hglue -0.7 true cm
\begin{tabularx}{15.5cm}{p{2.2cm}p{1.2cm}p{1.5cm}p{1.5cm}p{1.5cm}p{1.7cm}p{1cm}p{1.7cm}}
\hline \hline    
Experiment  & particles & beam energy {\small (GeV)} & $\sqrt{s}$ {\small (GeV)} &$x^\uparrow$  & $\cal L$ {\small(cm$^{-2}$s$^{-1}$)} & ${\cal P}_{\text{eff}} $  &  $\cal{F}$ (cm$^{-2}$s$^{-1}$) \\ 
\hline  
\AFTER b       		& $p+p^{\uparrow}$ 				& 7000     & 115 & $0.05\div 0.95$ & $1\cdot 10^{33}$   &  80\% & 6.4 $\cdot$ 10$^{32}$  \\
\AFTER b      		& $p+^{3}$He$^{\uparrow}$ 		& 7000     & 115 & $0.05\div 0.95$ & $2.5\cdot 10^{32}$    	&  23\% & 1.4 $\cdot$ 10$^{31}$  \\         
AFTER@ALICE$_\mu$      		& $p+p^{\uparrow}$ 				& 7000     & 115 & $0.1\div 0.3$ & $2.5\cdot 10^{31}$   &  80\% & 1.6 $\cdot$ 10$^{31}$  \\
\hline   
COMPASS (CERN)  	& $\pi^{\pm}+p^{\uparrow}$  	& 190      & 19  & $0.2\div 0.3$  & $2\cdot 10^{33}$ & 18\% & 6.5 $\cdot$ 10$^{31}$
   \\              
PHENIX/STAR (RHIC)	& $p^{\uparrow}+p^{\uparrow}$   & collider & 510 & $0.05\div 0.1$ & $2\cdot 10^{32}$ & 50\% &  5.0 $\cdot$ 10$^{31}$\\
E1039 (FNAL) 		& $p+p^{\uparrow}$       		& 120	   & 15  & $0.1\div 0.45$ & $4\cdot 10^{35}$ & 15\%  & 9.0 $\cdot$ 10$^{33}$ \\
E1027 (FNAL) 		& $p^{\uparrow}+p$       		& 120	   & 15  & $0.35\div 0.9$ & $2\cdot 10^{35}$ & 60\% & 7.2 $\cdot$ 10$^{34}$\\
NICA  (JINR) 		& $p^{\uparrow}+p$             	& collider & 26  & $0.1\div 0.8$  & $1\cdot 10^{32}$ & 70\%  & 4.9 $\cdot$ 10$^{31}$ \\
fsPHENIX (RHIC) 	& $p^{\uparrow}+p^{\uparrow}$	& collider & 200 & $0.1\div 0.5$  & $8\cdot 10^{31}$ & 60\%  & 2.9 $\cdot$ 10$^{31}$ \\
fsPHENIX (RHIC) 			& $p^{\uparrow}+p^{\uparrow}$	& collider & 510 & $0.05\div 0.6$ & $6\cdot 10^{32}$ & 50\% & 1.5 $\cdot$ 10$^{32}$ \\
PANDA (GSI)  		& $\bar{p}+p^{\uparrow}$       	&  15      & 5.5 & $0.2\div 0.4$  & $2\cdot 10^{32}$ & 20\% & 8.0 $\cdot$ 10$^{30}$  \\
\hline      \hline         
\end{tabularx}
\caption{
Compilation inspired from \cite{Barone:2010zz,Brodsky:2012vg} of the relevant parameters for the future or planned polarised DY experiments. The effective polarisation (${\cal P}_{\text{eff}} $ ) is a beam polarisation (where relevant) or an average polarisation times a (possible) dilution factor (for a gas target, similar to the one developed for HERMES~\cite{Steffens:2015kvp,0034-4885-66-11-R02,He3:pol:Milner}) or a target polarisation times a dilution factor (for the NH$_{3}$ target used by COMPASS and E1039). For \AFTER\, the numbers correspond to a gas target. $\cal{F}$ is the (instantaneous) spin figure of merit of the target defined as ${\cal F} ={\cal P}^{2}_{\text{eff}} \times {\cal L}$, with  ${\cal L}$ being the instantaneous luminosity.}
\label{tab:DY-SSA-projects}
\end{center}
\end{table}
As we mentioned in the previous section, Drell-Yan production is a unique probe of the Sivers effect for the quarks. It is a subject of lively interest with many existing or planned experiments (COMPASS, STAR, E1039). Table~\ref{tab:DY-SSA-projects} shows a compilation of the relevant parameters of future or planned polarised DY experiments. \AFTER\ is capable of measuring the Drell-Yan \An\ in a broad kinematic range with exceptional precision. 

\begin{figure}[hbt!]
	\centering
	
	\includegraphics[width=0.48\textwidth]{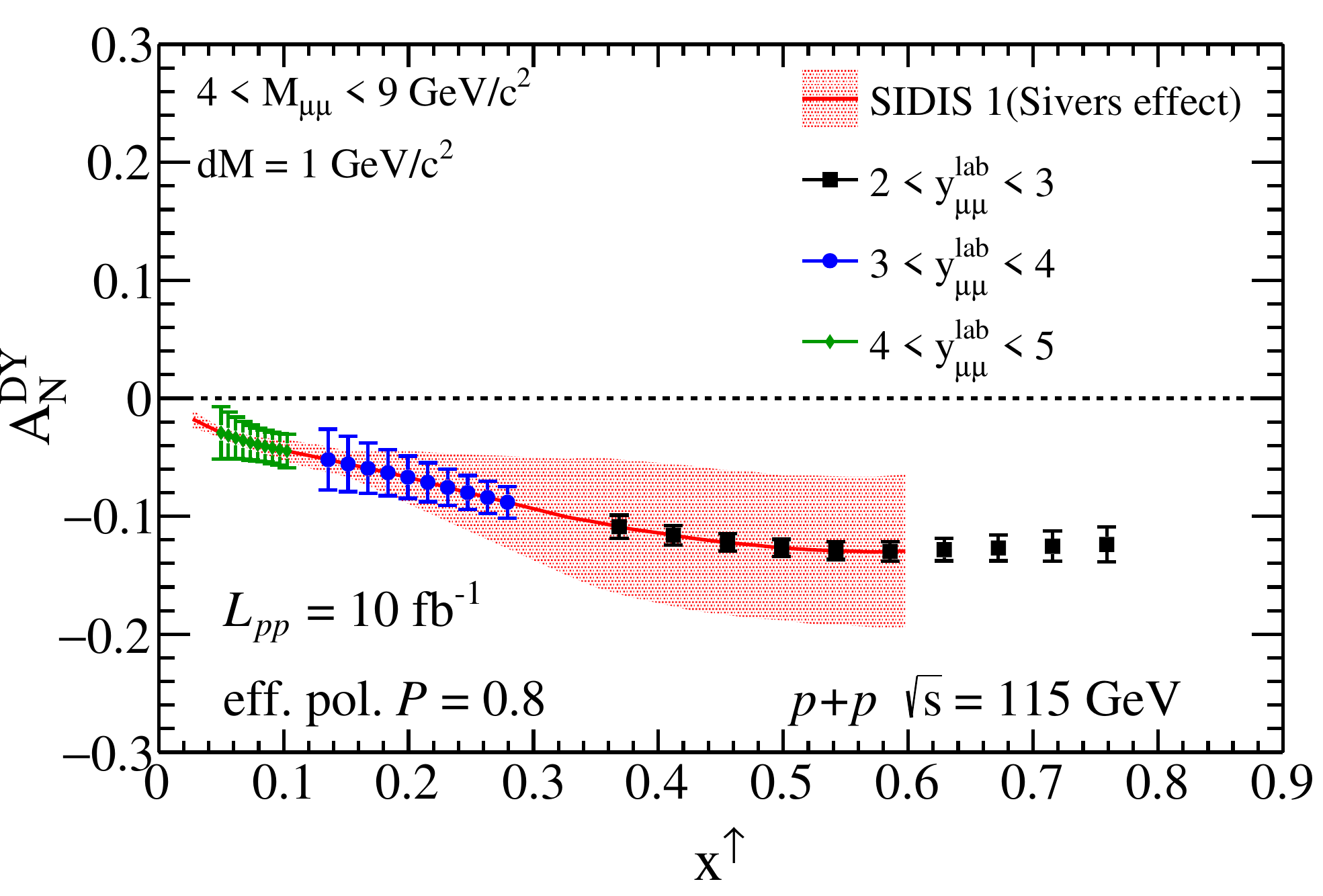}
	
	\caption[DY predictions]{Statistical projections for the Drell-Yan $\An$ measurement as a function of $x^{\uparrow}$ with a LHCb-like detector. Note that the range in $x^\uparrow$ is limited by the bin sizes in $y$ and $M$. We have checked that measurements can probably be done with an accuracy of $5\%$ up to $x^\uparrow \simeq 0.95$ as expected from Fig.~\ref{fig:mass:x2}.}
	\label{fig:An:Dy:LHCb}
\end{figure}

Figure~\ref{fig:An:Dy:LHCb} shows the statistical accuracy expected with a single data-taking year using a LHCb-like detector, for the Drell-Yan pairs satisfying $4< M_{\mu^+\mu^-} < 9 \, \gevcc$. The level of uncorrelated background drives the statistical uncertainties of this measurement. We estimated the background with the robust and commonly used like-sign technique, then we subtracted it from the $M_{\mu^+\mu^-}$ distribution. As explained above, we also assumed that the microvertexing detector allow one to remove the correlated background from charm and bottom pair decays. The \AFTER\ projections are compared to a theory evaluation~\cite{Anselmino:2015eoa}. This theory prediction based on SIDIS currently exhibit uncertainties much larger than our projected uncertainties, as shown by this example. By delivering high-quality data over a wide kinematic range, \AFTER\ will thus probe the $x^{\uparrow}$ dependence of the $A_{N}^{DY}$ and constrain model calculations.

Since the statistical precision of $A_{N}^{DY}$ strongly depends  on the level of uncorrelated background, such a study can be carried out with lower integrated luminosity if the background is suppressed.  The ALICE forward muon arm provides such a possibility. On the one hand, the available integrated luminosity is limited by the ALICE data taking rate capabilities. An integrated luminosity of $0.25$ fb$^{-1}$ can be expected at best for a single year of data taking. On the other, the absorber in front of its muon detector can potentially reduce the background. 

To check to which extent $A_{N}^{DY}$ can be studied with ALICE and if this deserves further in-depth investigations, we estimated the signal and background yields as follows. First, we assumed that the number of uncorrelated pairs in the DY and charmonium measurements is proportional to the number of charged particles squared, $N_{ch}^2$. To take into account the track reconstruction efficiency and the background suppression by the ALICE absorber, we referred to ALICE measurements of the di-muon pairs at forward rapidities in \pp\ collisions at 8~TeV~\cite{Adam:2015rta}. Since Drell-Yan pairs were not measured by ALICE yet, we turned to $J/\psi \rightarrow \mu^+ \mu^+$ measurements. We calculated the number of uncorrelated pairs $N_{un}$ under the \jpsi\ peak and the \jpsi\ yield and we normalised them by the luminosity. Then, we used the scaling of charged track density with the collision energy $dN_{ch}/d\eta|_{\eta \sim 0} = 0.725s^{0.23}$~\cite{Chatrchyan:2014qka} to scale down the $N_{un}$ to the level expected at $\sqrt{s} = 115$~GeV. Next, we used the energy dependence of the \jpsi\ cross section as calculated in the FONLL (Fixed Order plus Next-to-Leading Logarithms) framework to scale the \jpsi\ yield to the value expected at 115 GeV --admittedly this is another approximation. 
We neglected the change of the shape of $dN_{ch}/d\eta$ and $d\sigma^{J/\psi}/dy$ distributions with $\sqrt{s}$ for these first estimates. 
Finally, we assumed that the reconstruction efficiency for \jpsi\ and DY pairs was similar and the ratio of observed DY to \jpsi\ pairs would be the same in ALICE and LHCb. We thus used the correlated di-muon spectrum simulated for LHCb-like detector for $3 < y_{\mu^+\mu^-}^{\rm Lab} < 4$ and scale it to match the \jpsi\ yield in the simulation to that expected in ALICE at 115 GeV. Similarly, we normalised the uncorrelated background under the \jpsi\ peak in the simulations to the $N_{un}$ estimated for ALICE in a fixed target mode. As a results, we obtained distributions of DY, $c\overline{c} \rightarrow \mu^+\mu^-$, $b\overline{b} \rightarrow \mu^+\mu^-$ and uncorrelated background pairs expected in the measurements using the ALICE muon arm. 

\begin{figure}[hbt!]
	\centering
		\includegraphics[width=0.48\textwidth]{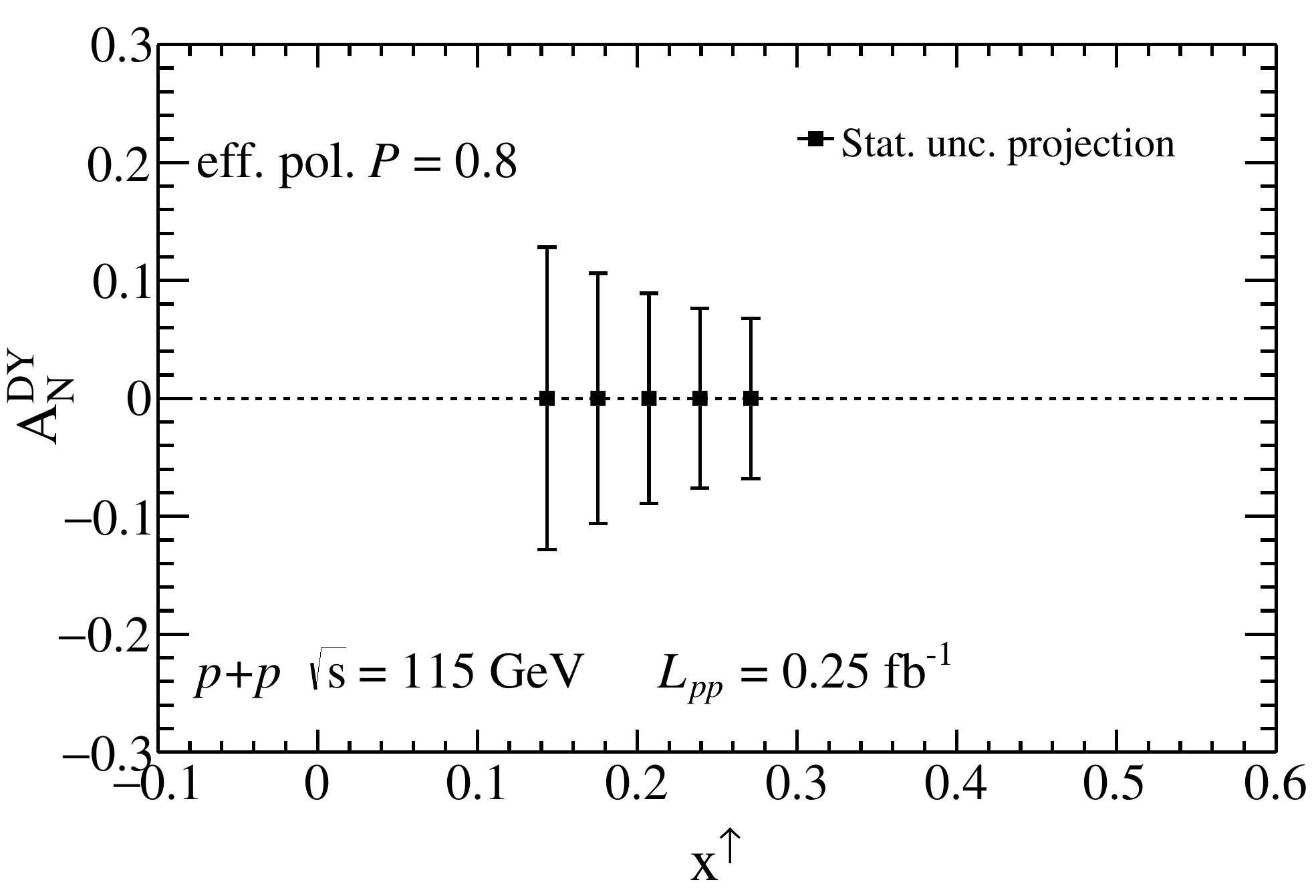} 
	\caption[Jpsi predictions]{Fast statistical projections for Drell-Yan $\An$ as a function of $x^{\uparrow}$ with a ALICE-like detector. See text for details.}
	\label{fig:An:DY:Alice}
\end{figure}

Note that we implicitly assumed the target location to be at $z = 0$, which is not completely coherent with the location of a polarised target. We nevertheless believe this to be sufficient for such a prospective study. 
Figure~\ref{fig:An:DY:Alice} shows the projections for \An\ measured with the ALICE-like acceptance of $3 < y_{\mu^+\mu^-}^{\rm Lab} < 4$. The uncertainties are sizeable but there is room for improvement if tracking and vertex detectors before the absorber are used to reject  muons from $\pi$, K meson and beauty and charm hadrons decays.

\subsection{Accessing the quark Sivers function in a polarised neutron: $p+^{3}$He$^{\uparrow}$ collisions.}

\begin{figure}[hbt!]
	\centering
	\includegraphics[width=0.48\textwidth]{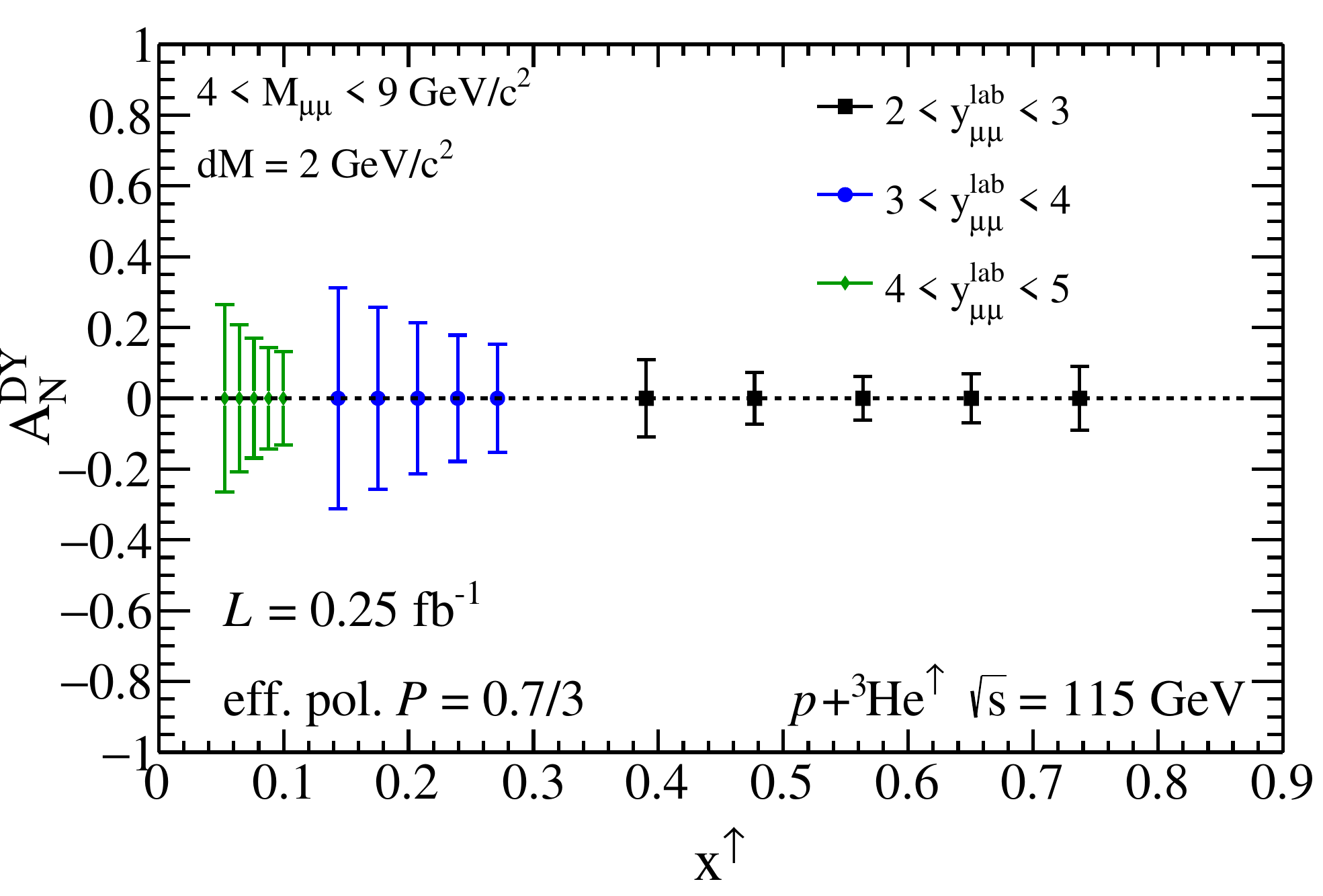} 
	\caption[Jpsi predictions]{Statistical projections for Drell-Yan $\An$ as a function of $x^{\uparrow}$ in $p+^{3}$He$^{\uparrow}$ collisions at $\sqrt{s}=115$~GeV. }
	\label{fig:An:DY:pHe}
\end{figure}

\begin{figure}[hbt!]
	\centering
	\begin{tabular}{cc}
		\includegraphics[width=0.48\textwidth]{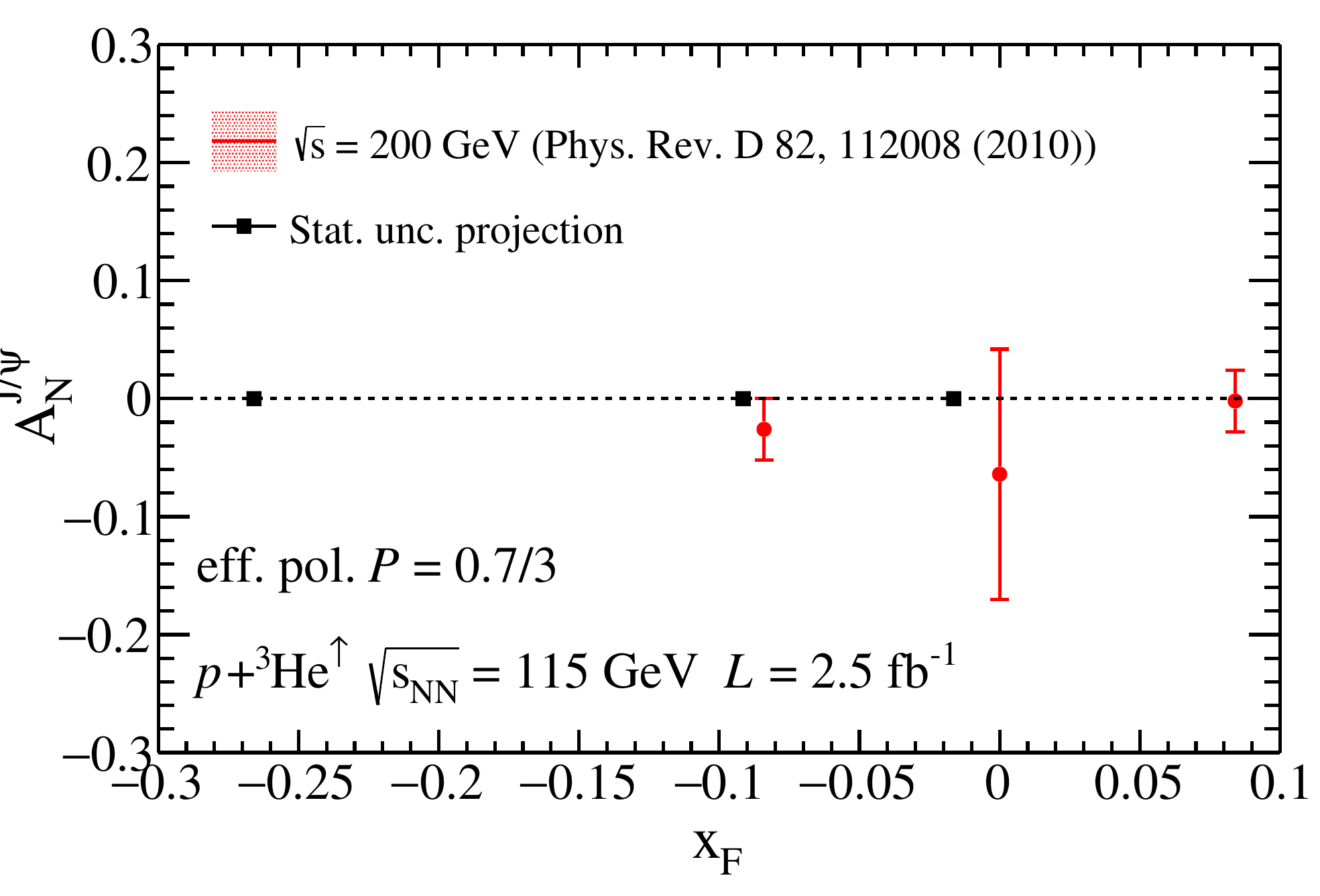} &
		\includegraphics[width=0.48\textwidth]{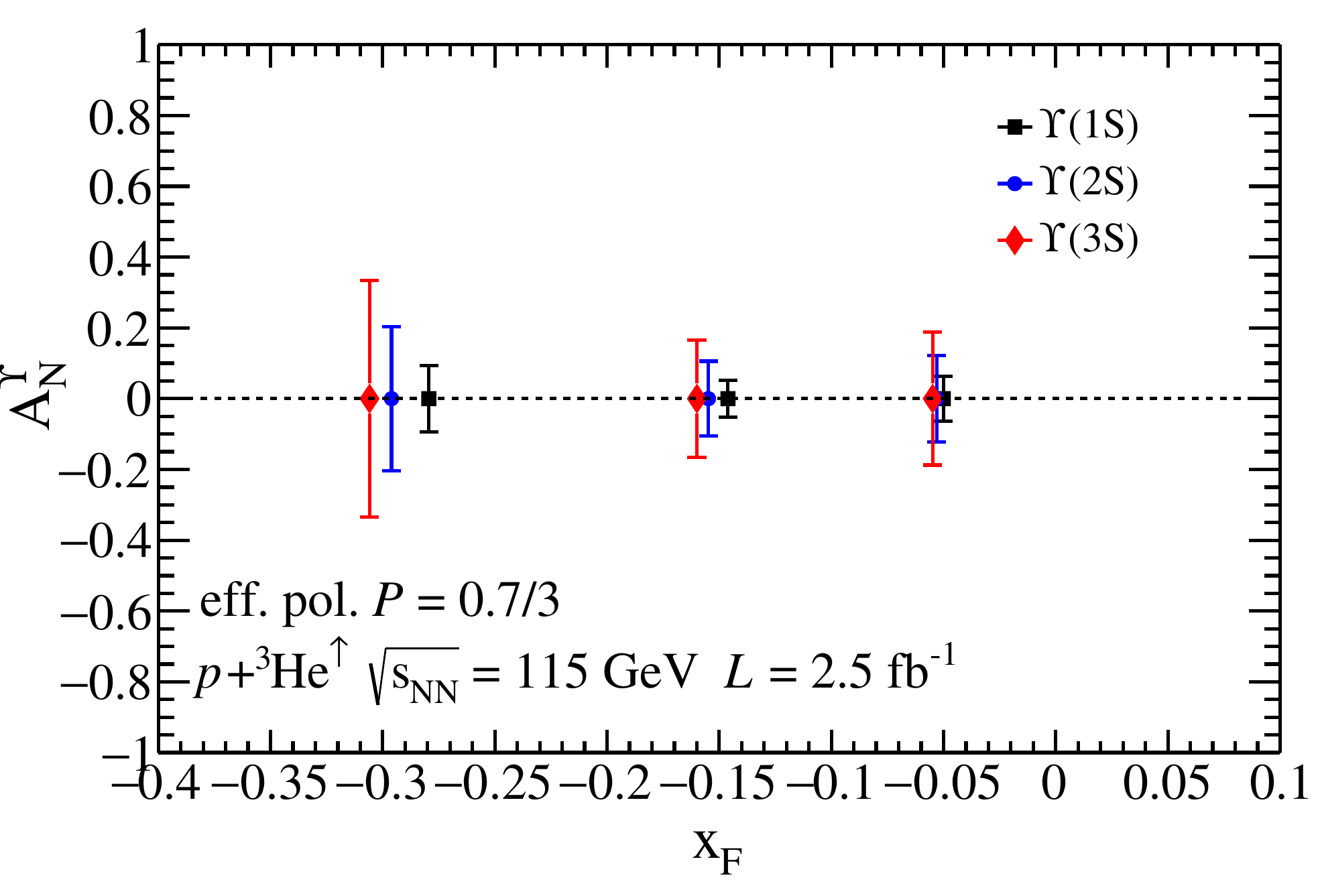} \\
	\end{tabular}
	\caption[Jpsi predictions]{Statistical projections for $\jpsi$ and \ups\ $\An$ as a function of $\xF$ in $p+^{3}$He$^{\uparrow}$ collisions at $\sqrt{s}=115$~GeV, compared to existing PHENIX data with polarised protons~\cite{Adare:2010bd}.}
	\label{fig:An:Quarkonium:pHe}
\end{figure}

\AFTER\ with a gas target offers a unique opportunity for studies of STSA in polarised $p+^{3}$He$^{\uparrow}$ collisions. Such reactions give access to polarised neutrons and thus to the Sivers functions in a neutron which can shed some light on its isospin dependence. Figures~\ref{fig:An:DY:pHe} and~\ref{fig:An:Quarkonium:pHe} show the statistical-uncertainty predictions for DY and quarkonium $A_{N}$ measurements. In the case of $^3$He$^\uparrow$, a polarisation of P = 70\% can be achieved~\cite{0034-4885-66-11-R02,He3:pol:Milner}. However, 
the effective polarisation, ${\cal P}_{\rm eff}$, is diluted by a factor of 3 since only the neutron is polarised in the $^3$He$^\uparrow$. 
In addition, the combinatorial background is proportional to the number of binary nucleon-nucleon collisions $N_{coll}$, thus the background increases by a factor $N_{coll} \approx \sqrt{3}$. An additional isospin factor of $9/6$ for DY studies is included. The available integrated luminosity of 2.5~fb$^{-1}$ will allow for an exploratory measurement for DY production and precision study for \jpsi\ $A_{N}$.

\section{Conclusions}
We have presented prospects and sensitivity studies for measurements of the 
quarkonia and Drell-Yan single transverse spin asymmetry which could be achieved
with a fixed-target experiment at the LHC, \AFTER, with a polarised gas target. 
Owing to its orginal acceptance, high target polarisation and large 
luminosities, \AFTER\ can deliver a set of unparalleled, high-quality data 
that will allow for in-depth studies of the gluon and quark Sivers functions and 
of the 3-parton collinear twist-3 correlators. 

The \An\ for \jpsi, \ups, C-even 
quarkonium states and various associated-production channels can be measured
 with unprecedented precision. The latter  will give access for the first time 
to the transverse momentum depedence of the gluon Sivers effect. 
Such studies, along with those of the STSAs for open charm and beauty,  
make \AFTER\ the best place to advance  --in a close future-- our knowledge 
of gluon and quark dynamics in a nucleon and of how they bind together to 
produce its observed spin $1/2$.

\section*{Acknowledgements}
This research was supported in part by the French P2IO Excellence Laboratory, the French CNRS via the grants FCPPL-Quarkonium4AFTER \&
TMD@NLO. AS acknowledges support from U.S. Department of Energy contract DE-AC05-06OR23177, under which Jefferson Science Associates, LLC, manages and operates Jefferson Lab.
MGE is supported by the Spanish Ministry of Economy and Competitiveness under the \emph{Juan de la Cierva} program and grant FPA2013-46570-C2-1-P.

\bibliographystyle{utphys}

\bibliography{AFTER-NewObservables-Spin}

\end{document}